\documentclass[runningheads]{llncs}

\usepackage[T1]{fontenc}
\usepackage{bm}
\usepackage[ruled,linesnumbered]{algorithm2e}
\usepackage{tablefootnote}
\usepackage{amsmath}
\usepackage{amssymb}
\usepackage{graphicx}
\usepackage{xcolor}

\begin{document}

\title{Factoring integers via \textit{Schnorr's algorithm} assisted with VQE}

\author{L. Sánchez Cano\inst{1}\orcidID{0009-0005-1017-1326} \and
G. Carrascal de las Heras\inst{2,3}\orcidID{0000-0001-7112-6696} \and
G. Botella Juan\inst{4,5}\orcidID{0000-0002-0848-2636} \and
A. del Barrio García\inst{6,7}\orcidID{0000-0002-6769-1200}}

\authorrunning{L. Sánchez Cano et al.}

\institute{Data Scientist, Iberdrola S.A., 28033 Madrid, Spain\\ 
\email{luisan13@ucm.es} \and 
IT Systems and Computation Department, Complutense University of Madrid, 28040 Madrid, Spain\\ 
\email{ginescar@ucm.es} \and 
IBM Quantum, IBM Consulting España, 28830 Madrid, Spain \and
Computer Architecture and Automation Department, Complutense University of Madrid, 28040 Madrid, Spain\\
\email{gbotella@ucm.es} \and 
Senior Member, IEEE
\and Computer Architecture and Automation Department, Complutense University of Madrid, 28040 Madrid, Spain\\
\email{abarriog@ucm.es} \and 
Senior Member, IEEE} 

\maketitle

\begin{abstract}
Current asymmetric cryptography is based on the principle that while classical computers can efficiently multiply large integers, the inverse operation, factorization, is significantly more complex. For sufficiently large integers, this factorization process can take in classical computers hundreds or even thousands of years to complete.
However, there exist some quantum algorithms that might be able to factor integers theoretically - the theory works properly, but the hardware requirements are far away from what we can build nowadays - and, for instance, Yan, B. et al.~(\cite{sublinearFact}) claim to have constructed a hybrid algorithm which could be able even to challenge \textit{RSA}-$2048$ in the near future. This work analyses this article and replicates the experiments they carried out, but with a different quantum method (VQE), being able to factor the number $1961$.

\keywords{Quantum computing \and Prime factorization \and VQE \and Cryptography \and RSA.}
\end{abstract}

\section{Introduction}

In this work we will discuss the possible breaking of RSA-2048 in the near future, based on a paper published in December, 2022~(\cite{sublinearFact}), where a hybrid algorithm is proposed to carry out the factorization of integers, specifically, a variant of \textit{Schnorr's algorithm}~(\cite{Schnorr91}). 
The objective of this paper is to replicate some of the experiments they conducted and propose a variant of \textit{Schnorr's algorithm} using the Variational Quantum Eigensolver (VQE). This approach aims to demonstrate an additional potential tool for addressing this topic.

The most time-consuming part of \textit{Schnorr's algorithm} is solving the Closest Vector Problem (CVP) and finding the smooth relation pairs (sr-pairs). This is where quantum computing comes into play. The quantum enhancement proposed by Yan, B. et al. in~\cite{sublinearFact} is based on the Quantum Approximate Optimization Algorithm (QAOA), and it would outperform \textit{Shor's algorithm}~(\cite{Shor}) in terms of the number of qubits required.

However, the experiments carried out in~\cite{Comment} to test this method have revealed some issues regarding its scalability, which raises questions about its practicality for larger-scale implementations.
 
 In addition, one of the main uncertainties is related to the time complexity of this algorithm~(\cite{sublinearFact}), making it hard to evaluate its potential. There also exist interesting results published in~\cite{Solano} using a non-hybrid algorithm (digitized-counterdiabatic quantum factorization), which actually works better than the one proposed by Yan, B. et al. It is a fact that interest in these applications is increasing more and more, being a current and relevant topic, as it can be seen in~\cite{QAforLatticeProbs}.
 
 The paper is organized as follows: \textbf{Section \(\ref{sec:schnorr}\)} provides details of \textit{Schnorr's algorithm} and includes comments on the work referenced in \cite{sublinearFact}. \textbf{Section \(\ref{sec:Quant}\)} discusses the quantum steps and proposes enhancements to \textit{Schnorr's algorithm}. Finally, \textbf{Section \(\ref{sec:exp}\)} describes the experiments we conducted and presents our conclusions.

\section{\textit{Schnorr's algorithm} and comments}\label{sec:schnorr}

First, we will briefly show \textit{Schnorr's algorithm} so as to discuss a variation of this algorithm later on, which is indeed the main topic of this paper.

This algorithm consists in the following:
\begin{itemize}
    \item Transform the factoring problem into a CVP (\textit{Closest Vector Problem)}.
    \item Get an approximate solution to the CVP using \textit{Babai's algorithm}.
    \item Obtain some smooth relation pairs.
    \item Solve a system of linear equations given by the previous pairs.
\end{itemize}

We can also see a flowchart containing \textit{Schnorr's algorithm} in figure \ref{fig:Flujo}. More details can be found in~\cite{SchnorrOriginalv2,Schnorr91,Schnorr2019}.

\begin{remark}\label{rema:SchnorrConj}

This however should be understood that due to the nature of the algorithm, it is based on an assumption and so there are issues with respect to how effective it is.

\end{remark}

\subsection{Construction of the CVP}

The first step of this algorithm is to convert the problem of factoring an integer $N=p\times q$ $(p<q)$ into a CVP on a lattice, which consists in, given a lattice and a target vector, finding the vector in that lattice closest to the target one. The reason behind this equivalence is a broad topic that can be seen in~\cite{Schnorr91}.

Consider a prime basis $P=\lbrace p_i\rbrace_{i=0,...,n}$, where $p_0=-1$ despite it is not a prime and $p_1,...,p_n$ are the first $n$ primes (all less than $p$). Let us define some concepts:

\begin{definition}
We say an integer $N=\displaystyle\prod_{i=1}^k w_i^{s_i}$ (where $w_1,...,w_k$ are its prime factors and $s_i \in \mathbb{N}\ \forall i$) is \bm{$p_n$}\textbf{-smooth} if $w_i\leq p_n$ $\forall i$. We also call this value of $n$ the \textbf{smooth bound} when $\lbrace p_i\rbrace_{i=0,...,n}$ are the first $n$ primes.
\end{definition}

\begin{definition}
An integer pair $(u,v)$ is called a \bm{$p_n$}\textbf{-smooth pair} if $u$ and $v$ are $p_n$-smooth. Moreover, such a pair is a \bm{$p_n$}\textbf{-smooth relation pair} (also called sr-pair) if
$$
u-vN=\displaystyle\prod_{i=0}^n p_i^{e_i} \text{ where } e_i \in \mathbb{N}.
$$
\end{definition}

Now, we construct the prime lattice $\Lambda(B_{n,c})$, where $c>0$ is an adjustable parameter. This is given by its matrix form $B_{n,c}=\left[b_1,...,b_n\right]\in \mathbb{R}^{(n+1)\times n}$, i.e.,
\begin{equation}
B_{n,c}=\begin{pmatrix}
    f(1) & 0 & \hdots & 0\\
    0 & f(2) & \hdots & 0\\
    \vdots & \vdots & \ddots & \vdots\\
    0 & 0 & \hdots & f(n)\\
    N^cln(p_1) & N^cln(p_2) & \hdots & N^cln(p_n)
\end{pmatrix},
\end{equation}

\noindent where $\left( f(1),...,f(n)\right)$ is a random permutation of the diagonal elements
$$\left( \sqrt{ln(p_1)},...,\sqrt{ln(p_n)}\right)$$
and $\lbrace b_1,...,b_n\rbrace$ is the lattice basis (where $\lbrace b_i\rbrace_{i=1}^n$ are linearly independent vectors). Additionally, we construct the target vector $t\in \mathbb{R}^{n+1}$ as 
\begin{equation}
t=\begin{pmatrix}
    0\\
    \vdots\\
    0\\
    N^cln(N)
\end{pmatrix}.
\end{equation}

\noindent However, it is important to remark that in~\cite{sublinearFact}, when it comes to doing calculations, they propose the following way to create $B_{n,c}$ and $t$ instead of the previous construction (which was a proposal by Schnorr):
\begin{equation}
B_{n,c} = \begin{pmatrix}
    f(1) & 0 & \hdots & 0\\
    0 & f(2) & \hdots & 0\\
    \vdots & \vdots & \ddots & \vdots\\
    0 & 0 & \hdots & f(n)\\
    \lfloor 10^cln(p_1)\rceil & \lfloor 10^cln(p_2)\rceil & \hdots & \lfloor 10^cln(p_n)\rceil
\end{pmatrix}\ t = \begin{pmatrix}
    0\\
    \vdots\\
    0\\
    \lfloor 10^cln(N)\rceil
\end{pmatrix},
\end{equation}

\noindent where $\lbrace f(i)\rbrace_{i=1}^n$ is a random permutation of $\lbrace \lfloor \frac{i}{2} \rceil \rbrace_{i=1}^n$.

It is relevant to mention that $n$ (dimension of the lattice) corresponds to the number of qubits we will need and such magnitude is given by 
\begin{equation}
n = \lfloor \frac{l * log_2(N)}{log_2(log_2(N))}\rfloor,
\end{equation} 

\noindent where $l \in \lbrace 1,2\rbrace$ is a hyperparameter (see~\cite{Comment}), which will be considered to be $1$ unless otherwise stated. Besides this, the latter construction will be used in our experiments too in order to compare them to those in~\cite{sublinearFact}.

The next goal is to find a vector $b_0\in\Lambda(B_{n,c})$ which is the closest one to our target vector $t$, which means $b_0 = arg\ \underset{b\in\Lambda}{min}\lVert b-t\rVert$. In order to find an approximate solution to this, the following step is to use \textit{Babai's algorithm}.

\subsection{LLL-reduction}

Before addressing \textit{Babai's algorithm}, let us focus on one of the major components of this algorithm which is called the LLL-reduction algorithm (for further details see~\cite{LLLVar} and~\cite{LLLred}). 
This is a polynomial time lattice basis reduction algorithm.

First, let us recall the Gram-Schmidt (GS) orthogonalization process. Given a basis $\lbrace b_1,...,b_n\rbrace$, its Gram-Schmidt basis is given by 
\begin{equation}
\tilde{b}_i=b_i-\displaystyle\sum_{j=1}^{i-1}\mu_{ij}\tilde{b}_j,
\end{equation}

\noindent where $\mu_{ij}=\frac{\langle b_i,\tilde{b}_j\rangle}{\langle \tilde{b}_j,\tilde{b}_j\rangle}$ are the GS coefficients. This new basis is an orthogonal basis, but the vectors are not necessarily in the lattice.

We say a lattice basis $\lbrace b_1,...,b_n\rbrace$ is \textbf{$\delta$-LLL-reduced}, where $\delta\in \left(\frac{1}{4},1\right]$, if it satisfies the following:
\begin{enumerate}
    \item \textbf{Size-reduced}: $|\mu_{ij}|\leq \frac{1}{2}$ for $1\leq j < i \leq n$ (this property guarantees the length reduction of the ordered basis).
    \item \textbf{Lovász condition}: $\lVert \tilde{b}_i+\mu_{i,i-1}\tilde{b}_{i-1}\rVert^2\geq \delta\lVert\tilde{b}_{i-1}\rVert^2$ for $1<i\leq n$.
\end{enumerate}

\begin{remark}
Although LLL-reduction is well-defined for $\delta =1$, the polynomial-time complexity is guaranteed only for $\delta  \in (\frac{1}{4}, 1)$.
\end{remark}

We tried to reproduce the pseudocode of LLL-reduction given in~\cite{sublinearFact}, using the $3$-qubit case they show. However, the results we obtained were different from those presented in this work, that is why we tried to figure out what pseudocode they used. The input is
\begin{equation}\label{eqn:input3q}
    B = \begin{pmatrix}
        1 & 0 & 0\\
        0 & 1 & 0\\
        0 & 0 & 2\\
        22 & 35 & 51
    \end{pmatrix},\ \delta=\frac{3}{4}
\end{equation}

\noindent and we obtain the output
\begin{equation}
    B = \begin{pmatrix}
        1 & -2 & -1\\
        0 & 1 & -1\\
        0 & 0 & 2\\
        22 & -9 & -6
    \end{pmatrix},
\end{equation}

\noindent whereas the result in~\cite{sublinearFact} is 
\begin{equation}\label{eqn:result3q}
    B = \begin{pmatrix}
        1 & -4 & -3\\
        -2 & 1 & 2\\
        2 & 2 & 0\\
        3 & -2 & 4
    \end{pmatrix}
\end{equation}

\noindent The subsequent method we proceeded was he LLL-reduction algorithm given in Algorithm \ref{alg:LLL}. 
\begin{algorithm}
  \caption{LLL-reduction algorithm}
  \label{alg:LLL}
  \KwData{Lattice basis $B=\lbrace b_i\rbrace_{i=1}^n \subset \mathbb{R}^{n+1}$, parameter $\delta \in \left(\frac{1}{4}, 1\right]$ }
  \KwResult{$\delta$-LLL-reduced basis}
  
  GS orthogonalization: we get a new basis $\tilde{B}=\lbrace \tilde{b}_i\rbrace_{i=1}^n$ and the coefficients $\mu_{i,j}\ \forall i,j$.\\Set $k=2$.\\
 \While{$k\leq n$}{\For{$j$ from $k-1$ to $1$}{\If{$|\mu_{k,j}|>1/2$}{$b_k=b_k-\lfloor\mu_{k,j}\rceil b_j$\\ Update $\tilde{B}$ and the coefficients $\mu_{r,s}\ \forall r,s$} } \eIf{$\langle \tilde{b}_k,\tilde{b}_k\rangle\geq (\delta-\mu_{k,k-1}^2)\langle \tilde{b}_{k-1}, \tilde{b}_{k-1}\rangle$}{$k=k+1$}{Swap the order of $b_k$ and $b_{k-1}$\\ Update $\tilde{B}$ and the coefficients $\mu_{r,s}\ \forall r,s$\\ $k=max(k-1,2)$} }
  \Return{$\delta$-LLL-reduced basis $B$}
\end{algorithm}

\noindent This algorithm is slightly different from the original LLL-reduction algorithm (presented in~\cite{LLLred}), since we have considered a general $\delta\in\left(\frac{1}{4},1\right]$, but it was first shown for $\delta=\frac{3}{4}$. In addition, for each $k$ they~(\cite{LLLred}) only check size-reduction for $j = k-1$ before verifying Lovász condition, and afterwards they continue with the size-reduction for that $k$ and once size-reduction is finished, $k$ is incremented. However, we have followed the main idea from~\cite{LLLVar} and~\cite{SchnorrLLL} so as to construct this algorithm.

Thus, if we use Algorithm \ref{alg:LLL} with the previously mentioned input (equation \ref{eqn:input3q}), we obtain
\begin{equation}
B = \begin{pmatrix}
        1 & -3 & -4\\
        -2 & 2 & 1\\
        2 & 0 & 2\\
        3 & 4 & -2
    \end{pmatrix},
\end{equation}

\noindent which is slightly different from the result presented in~\cite{sublinearFact} (equation \ref{eqn:result3q}). Yet, our result is a $\delta$-LLL-reduced basis too.

The distinction concerning~\cite{sublinearFact} after performing algorithm \ref{alg:LLL} is due to the possible use of a different algorithm known as the $\bm{L^2}$\textbf{ algorithm} (LLL-reduction algorithm in floating point), which was described in~\cite{L2red}. The explanation for this is that the authors receive the output produced by a Python module named \textit{fpylll}, which is based on the L2 algorithm.

\subsection{\textit{Babai's algorithm}}

This algorithm (also called Babai's nearest plane algorithm) is useful to get an approximate solution to CVP. Indeed, when we have a LLL-reduced basis $B=\lbrace b_1,...,b_n\rbrace$, the solution $b_{op}$ given by this algorithm satisfies 
\begin{equation}
\lVert t-b_{op}\rVert \leq 2^{n/2}\lVert t - b_0\rVert, 
\end{equation}

\noindent where $t$ is the target vector and $b_0$ is the closest vector to $t$ in the lattice. The details of Babai's algorithm can be found in \cite{Babai}. 

Let us reproduce the pseudocode shown in~\cite{sublinearFact} (considering the LLL-reduction step with the \textit{fpylll} module) to see what we get. The input in this case is the one given by Equation \ref{eqn:input3q} and also the target vector
\begin{equation}
    t = \begin{pmatrix}
        0\\
        0\\
        0\\
        240
    \end{pmatrix}.
\end{equation}

\noindent The output we obtain is
\begin{equation}
    b_{op} = \begin{pmatrix}
        877\\
        -110\\
        -66\\
        14001
    \end{pmatrix},
\end{equation}

\noindent whereas their outcome is
\begin{equation}
    b_{op} = \begin{pmatrix}
        0\\
        4\\
        4\\
        242
    \end{pmatrix}.
\end{equation}

\noindent Afterwards, we opted for the \textit{Babai's algorithm} described in the \textit{fpylll} module as the LLL-reduction algorithm performed successfully with it and the results we obtained from this algorithm are in line with those in \cite{sublinearFact}. In conclusion, there exists reasonable evidence to think this \textit{fpylll} module could be the one behind their experiments.

\subsection{Obtaining sr-pairs}\label{subsec:sr-pairs}

Now, we will see how given a point in the lattice we can obtain a $p_n$-smooth pair. Let $b=\displaystyle\sum_{i=1}^n e_ib_i\in\Lambda(B_{n,c})$, where $e_i\in\mathbb{Z}\ \forall i$. We can consider
\begin{equation}
u=\displaystyle\prod_{e_i>0} p_i^{e_i} \text{ and } v=\displaystyle\prod_{e_i<0}p_i^{-e_i},
\end{equation}

\noindent and it can be seen they are both $p_n$-smooth. Hence, we have shown that a vector on a lattice encodes a $p_n$-smooth pair, denoted $b=(e_1,...,e_n)\sim (u,v)$.

We expect to obtain the sought sr-pairs among the $(u,v)$-pairs given by the lattice vectors we get when trying to solve the CVP.

\subsection{Solving the system of linear equations}

Once we have one or more sr-pairs, for instance,

\begin{equation}
    \begin{cases}
      (u_1, v_1) \sim (e_{11}, ..., e_{n1})\\
      (u_2, v_2) \sim (e_{12}, ..., e_{n2})\\
      \ \ \vdots\\
      (u_k, v_k) \sim (e_{1k}, ..., e_{nk})
    \end{cases}\,,
\end{equation}

\noindent we just need to solve the following system of linear equations to try to get the factors: 

\begin{equation}
    \begin{cases}
     e_{11}t_1 + ... + e_{1k}t_k \equiv 0 \text{ mod } 2\\
     e_{21}t_1 + ... + e_{2k}t_k \equiv 0 \text{ mod } 2\\
     \ \ \vdots\\
     e_{n1}t_1 + ... + e_{nk}t_k \equiv 0 \text{ mod } 2
    \end{cases}\,
\end{equation}

\section{Quantum steps}\label{sec:Quant}

Given that solving a system of linear equations is a polynomial-time process, it can be argued that factoring \(N\) is computationally equivalent to identifying sr-pairs. This equivalence highlights why quantum computing becomes relevant between the implementation of \textit{Babai's algorithm} and the part we outline in Section \ref{subsec:sr-pairs}, the obtaining of sr-pairs.

\subsection{Problem Hamiltonian}

The main goal of this part is to improve the solution given by \textit{Babai's algorithm}, let us explain how.  

Let $B=\lbrace b_1,...,b_n\rbrace$ be an LLL-reduced basis of the lattice $\Lambda$ and $b_{op} \in \Lambda$ be the solution given by \textit{Babai's algorithm}. To construct the Hamiltonian we will consider the Euclidean distance between the target vector $t$ and a new vector 
\begin{equation}
b_h = b_{op} + \displaystyle\sum_{i=1}^n x_i b_i,   
\end{equation}

\noindent where $x_i \in \lbrace 0,1\rbrace$. In other words, once we have reached a sufficiently ``good'' region (where $b_{op}$ is), we move around that region a little bit in the direction of the basis, while remaining inside the lattice. Hence, the cost function of the problem is 
\begin{equation}
F(x_1,...,x_n) = \lVert t-b_h\rVert^2 = \lVert t - b_{op} - \displaystyle\sum_{i=1}^n x_i b_i\rVert^2, 
\end{equation}

\noindent i.e., it is a QUBO (\textit{Quadratic Unconstrained Binary Optimization}) problem. However, we would like to transform it into an Ising Hamiltonian, where variables belong to $\lbrace 1, -1\rbrace$. In order to do so, we consider $x_i = \frac{z_i+1}{2}$, with $z_i \in \lbrace 1, -1\rbrace$.

\begin{remark}
It is easy to see that for $z_i = 1$, we have $x_i = 1$, and for $z_i = -1$,  we have $x_i = 0$.
\end{remark}

Therefore, for our purpose, the Hamiltonian is given by 
\begin{equation}
\hat{H} = \lVert t - b_{op} -\displaystyle\sum_{i=1}^n\hat{x}_i b_i\rVert^2,
\end{equation}

\noindent where $\hat{x}_i = \frac{\sigma_z^i+I}{2}$ is a quantum operator, %$c_i = \lfloor \mu_i \rceil$, 
$I$ is the $2\times 2$ identity matrix and $\sigma_z^i$ represents the tensor product of $n$ identity matrices, except for the matrix in position $i$, which is the Pauli-$z$ matrix $\sigma_z  = \begin{pmatrix}
        1 & 0\\
        0 & -1
\end{pmatrix}$, i.e., $\sigma_z$ is applied to the qubit in position $i$, whereas the rest of the qubits remain the same. Therefore, we get a $2^n \times 2^n$ Hamiltonian.

\subsection{Quantum enhancement}

After some of the results mentioned in the introduction were published, we decided to propose another possible approach: \textit{Schnorr's algorithm} plus Variational Quantum Eigensolver. The classical steps are the ones we previously mentioned, but there are changes in the quantum part with respect to~\cite{sublinearFact}.

\begin{figure}[h]
\centering
\includegraphics[width=0.65\linewidth]{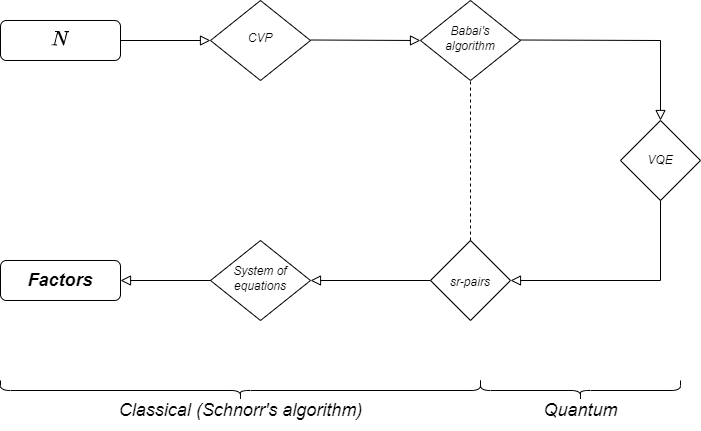}
\caption{\label{fig:Flujo} Flowchart of the proposed algorithm. The difference with respect to \textit{Schnorr's algorithm} is the addition of VQE. Besides this, with respect to~\cite{sublinearFact}, QAOA has been replaced by VQE.}
\end{figure}

Let us display a flowchart of the algorithm we propose (see Figure \ref{fig:Flujo}).

Here (Figure \ref{fig:Flujo}) we see that the goal of VQE is to improve the approximate solution given by \textit{Babai's algorithm} in order to find sufficiently good sr-pairs.

QAOA may converge faster, but when implemented on current circuits, it faces challenges. With VQE, on the other hand, we have shorter circuits which are therefore less susceptible to noise in this specific case (a discussion on this can be seen in~\cite{carrascal2023backtesting}). Moreover, regarding Babai's solution, QAOA can take advantage of it as an \textit{ansatz}, but VQE does not need it actually. Nevertheless, such solution is useful regardless of the method we use since it is utilized to construct the Hamiltonian $\hat{H}$.

\section{Experiments}\label{sec:exp}

\subsection{Results}

Let us display the results we have obtained using Python code on Google Colab utilizing frameworks like \textit{Qiskit} and the module \textit{fpylll} based on the methodology we have shown in preceding sections. We will show the process for $N = 1961$, $l = 1$, $c = 1.5$, and $15$ as \textit{smooth bound}. During these calculations, we will use \textit{fpylll} module for LLL-reduction and \textit{Babai's algorithm}.

First, we get that the number of qubits needed is $n = 3$, so the prime basis is $P = \lbrace -1, 2, 3, 5\rbrace$. In this case, the matrix of the lattice and the target vector are 
\begin{equation}
    B = \begin{pmatrix}
        1 & 0 & 0\\
        0 & 1 & 0\\
        0 & 0 & 2\\
        22 & 35 & 51
    \end{pmatrix},\ t = \begin{pmatrix}
        0\\
        0\\
        0\\
        240
    \end{pmatrix}.
\end{equation}

\noindent Next, we carry out the LLL-reduction of $B$, obtaining
\begin{equation}
    B = \begin{pmatrix}
        1 & -4 & -3\\
        -2 & 1 & 2\\
        2 & 2 & 0\\
        3 & -2 & 4
    \end{pmatrix}.
\end{equation}

\noindent Afterwards, applying \textit{Babai's algorithm} with this LLL-reduced basis, we get the approximate solution
\begin{equation}
    b_{op} = \begin{pmatrix}
        0\\
        4\\
        4\\
        242
    \end{pmatrix}.
\end{equation}

\noindent It is here where quantum computing arises in order to find the minimum for the Hamiltonian. We can see the eigenvectors and eigenvalues given by VQE in table \ref{table:VQE}.

\begin{table}
\centering
\caption{Results of VQE for the $3$-qubit case.}\label{table:VQE}
\begin{tabular}{| c | c | c |}
\hline 
\textit{Selection} & \textit{Value} & \textit{Probability}\\ \hline
$\left[ 0 0 0 \right] $ & 36 & 1.0 \\ \hline
$\left[ 1 0 0 \right] $ & 66 & 0.0 \\ \hline
$\left[ 0 0 1 \right] $ & 97 & 0.0 \\ \hline
$\left[ 1 1 0 \right] $ & 91 & 0.0 \\ \hline
$\left[ 1 1 1 \right] $ & 174 & 0.0 \\ \hline
$\left[ 0 1 1 \right] $ & 150 & 0.0 \\ \hline
$\left[ 0 1 0 \right] $ & 77 & 0.0 \\ \hline
$\left[ 1 0 1 \right] $ & 137 & 0.0 \\ \hline
\end{tabular}
\end{table}

\noindent Therefore, the optimal selection is $\left[000\right]$, whose value is $36$. Once we have this, the next step is to find sr-pairs. We have based our implementation of this step on the work~\cite{Comment}, obtaining the sr-pair $\left(2025, 1\right)$, which in fact corresponds to the $4$-th sr-pair in~\cite{sublinearFact}. Afterwards, we solve the system of equations given by that sr-pair, finally calculating the factors: $53$ and $37$.

This means our code has worked properly, since $53 \times 37 = 1961$, i.e., we have factored the given number using VQE in a similar way as they did with QAOA in~\cite{sublinearFact}.

\begin{remark}
This experiment has been carried out considering the same diagonal order $\left[1, 1, 2\right]$ for $B$ as in \cite{sublinearFact}. Remember this diagonal is supposed to be a random permutation of some elements.
\end{remark}

We have also replicated the experiment considering the $\delta$-LLL-reduced basis we calculated with algorithm \ref{alg:LLL} instead of the one given by the \textit{fpylll} module. However, we ended up with no sr-pairs found, so we could not find the factors of $N=1961$ in this case.

Besides, we have tried to replicate the experiment for the 5-qubit case ($N=48567227$) with both ways to calculate the $\delta$-LLL-reduced basis, but in these two attempts we did not get any sr-pair. Lastly, we have carried out all these experiments once more, but considering the hyperparameter $l=2$. To sum up, the results are indicated in table \ref{table:Results}.

\begin{table}
\centering
\caption{Results of experiments.}\label{table:Results}
\begin{tabular}{| c | c | c | c | c | c | c | c | c |}
\hline 
$N$ & $l$ & \textit{qubits} & $c$ & $LLL$\textit{-reduction} & \textit{SB}\tablefootnote{\textit{SB} stands for \textit{smooth bound}}& \textit{uv-pairs} & \textit{sr-pairs} & \textit{factors} \\ \hline
1961 & 1 & 3 & 1.5 & \textit{fpylll} & $15$ & $(2025, 1)$ & $(2025, 1)$ & $\boldsymbol{53, 37}$\\ \hline
1961 & 1 & 3 & 1.5 & algorithm \ref{alg:LLL} & $15$ & $(30375, 16384)$ & none & \textbf{failed}\\ \hline
1961 & 2 & 6 & 1.5 & \textit{fpylll} & $15$ & $(1950, 1)$ & $(1950, 1)$ & \textbf{failed}\\ \hline
1961 & 2 & 6 & 1.5 & algorithm \ref{alg:LLL} & $15$ & $(999635, 1185921)$ & none & \textbf{failed}\\ \hline
48567227 & 1 & 5 & 4 & \textit{fpylll} & $50$ & $(48620250, 1)$ & none & \textbf{failed}\\ \hline
48567227 & 1 & 5 & 4 & algorithm \ref{alg:LLL} & $50$ & $(311..., 313...)$\tablefootnote{$(311402196875, 313456656384)$} & none & \textbf{failed}\\ \hline
48567227 & 2 & 10 & 4 & \textit{fpylll} & $50$ & $(48586629, 1)$ & $(48586629, 1)$ & \textbf{failed} \\ \hline
48567227 & 2 & 10 & 4 & algorithm \ref{alg:LLL} & $50$ & \textbf{failed}\tablefootnote{Failure due to the limit of compile time} & \textbf{failed} & \textbf{failed}\\ \hline
\end{tabular}
\end{table}

\noindent Here we see that most of the experiments ($7$ out of $8$) failed. This could be due to the use of a seed to replicate them, i.e., we have just run each of them once. Besides this, in contrast to QAOA, VQE only gives one possible solution to the CVP, leading to lower probability of success in each execution. Furthermore, what was mentioned in Remark \ref{rema:SchnorrConj} regarding \textit{Schnorr's algorithm}  could have also contributed to the failures.

It is worth mentioning that in the experiments where $l = 2$, we have not considered the diagonal elements of $B$ to be in the same order as in \cite{sublinearFact} since there are no such examples in that work.

\subsection{Conclusions}

We propose a methodology for factoring integers using \textit{Schnorr's algorithm} and employing the Variational Quantum Eigensolver (VQE). This approach has yielded successful results in a sample case. While the proposed algorithm is not flawless, with sufficient iterations and appropriate adjustments to hyperparameters, such as the number of qubits, it holds the potential to factorize the targeted integer in specific scenarios.

Moreover, it has been noted that increasing the number of qubits does not necessarily improve outcomes. Adding more qubits doubles the dimensionality and might enhance accuracy, but it also increases the risk of the algorithm becoming stuck in local minima. Crucially, even a single successful attempt at factoring integers can have profound and potentially disastrous consequences, especially in critical areas like bank transfers, secure communications, and national security, where a single breach could compromise sensitive data or undermine critical infrastructure.

In conclusion, delving deeper into this topic seems both logical and promising. It would be interesting to execute these experiments on real NISQ (Noisy Intermediate-Scale Quantum) hardware of a more expansive scale, even implementing hybrid quantum-classical solutions with High-Performance Computing. Furthermore, as highlighted in \cite{sublinearFact}, there is an estimation that their algorithm could challenge \textit{RSA}-$2048$ with 372 physical qubits. This provides an additional compelling reason to harness the capabilities of real NISQ machines. 

\begin{credits}
\subsubsection{\ackname} This work was supported by grant PID2021-123041OB-I00 funded by MCIN/AEI/ 10.13039/501100011033 and by “ERDF A way of making Europe”. We acknowledge the use of IBM Quantum services for this work. The views expressed are those of the authors and do not reflect the official policy or position of IBM or the IBM Quantum team.
\subsubsection{\discintname}
We declare that the authors have no competing interests. 
\end{credits}

\end{document}